\documentclass[12pt,preprint]{aastex}

\newcommand{\CII}{C\,{\sc ii}}
\newcommand{\Viv}{V\,{\sc iv}}
\newcommand{\Viii}{V\,{\sc iii}}
\newcommand{\Tiiii}{Ti\,{\sc iii}}
\newcommand{\Hi}{H\,{\sc i }}
\newcommand{\Hii}{H\,{\sc ii }}
\newcommand{\Tiii}{Ti\,{\sc ii}}

\newcommand{\Vii}{V\,{\sc ii}}
\def\Msun{{\hbox {M$_\odot$}}}

\slugcomment{}

\shorttitle{New UfIB in NGC~7027}
\shortauthors{Goicoechea et al.}

\begin{document}

\title{A new Unidentified Far Infrared Band in NGC~7027}

\author{Javier R. Goicoechea, Jos\'e Cernicharo, Helena
Masso and Mar\'{\i}a Luisa Senent}
\affil{Departamento de Astrof\'{\i}sica Molecular e Infrarroja, 
Instituto de Estructura de la Materia, CSIC, 
Serrano 121, E--28006 Madrid, Spain.} 
\email{javier@damir.iem.csic.es, cerni@damir.iem.csic.es,
masso@damir.iem.csic.es,
imts420@iem.cfmac.csic.es}

\begin{abstract}

We report on  the detection of a molecular band centered at $\sim$98~$\mu$m
($\sim$102~cm$^{-1}$), observed with the $\textit
{Infrared Space Observatory}$$^1$
in the young Planetary Nebula NGC~7027. The band structure and intensity
can not be reproduced by atomic fine structure lines, recombination
lines or by
the rotational emission of abundant molecules.
We discuss the possible contribution of the
low--energy bending modes of pure carbon chains to the
unidentified far--IR bands observed in  $C$--rich evolved objects.
In particular, we speculate that the band emission 
could arise from the 
$\nu_9$ and $\nu_7$ bending modes of C$_6$ and C$_5$, for
which wavenumbers of 90$\pm$50 and  107$\pm$5~cm$^{-1}$ have been
estimated from photoelectron spectroscopy.

\end{abstract}


\footnotetext[1]
{Based on observations with ISO,
an ESA project with instruments funded by ESA Member States
(especially the PI countries: France, Germany, the Netherlands
and the United Kingdom) and with participation of ISAS and NASA.}

\section{Introduction}

Carbon chains molecules are of special interest for Astrophysics
since Douglas (1977) proposed them as possible carriers
of the diffuse interstellar bands (DIBs).
In the past years, many polar carbon chains have been detected in
the interstellar and circumstellar media (ISM and CSM)
through their pure rotational spectrum at radio wavelengths.
Major examples are the cyanopolyynes, HC$_{2n+1}$N with $n$=1--5
(Turner 1971, Avery et al. 1976, Kroto et al. 1978, Broten et al. 1978,
Bell et al. 1997), and the hydrogenated carbon chain radicals such
as C$_5$H, C$_6$H, C$_7$H  and C$_8$H (Cernicharo et al. 1986,
Gu\'elin et al. 1987, 1997, Cernicharo \& Gu\'elin 1996).

On the other hand, larger carbon molecular complexes such as the
polyatomic aromatic hydrocarbons (PAHs) are believed to
dominate the ubiquitous mid--IR emission seen in the  unidentified
infrared bands (UIBs). However, only the single aromatic species,
benzene, has been so far identified through its IR active modes
(Cernicharo et al. 2001).

All this chemistry richness suggests that the growth mechanisms
producing complex carbon molecules such as the PAHs or the fullerenes
are highly efficient. Nevertheless, the set of `building blocks''
and the possible chemical reactions still have to be identified and
clarified.
Among the possible species, the pure carbon chains,
(C$_n$; Van Orden \& Saykally 1998), could be
the `skeletons'' from which larger organic molecules can be formed
(see Cernicharo, Goicoechea \& Caux 2000, Cernicharo 2004).
Due to the lack of permanent electric dipole, these species
do not have rotational spectrum to be observed from radio telescopes.
The only way to detect them
in the dense ISM and CSM is through their asymmetrical stretching
modes around $\sim$5~$\mu$m (2000~cm$^{-1}$) and/or through their
low--energy bending modes around
$\sim$100~$\mu$m ($\sim$100~cm$^{-1}$).
The first technique  allowed the detection of C$_3$ (Hinkle et al.
1988) and C$_5$ (Bernath et al. 1989) in the circumstellar
envelope of the bright IR evolved star IRC+10216.
However, few sources (specially in the ISM) have
enough flux at $\sim$5~$\mu$m to allow systematic studies of
the C$_n$ stretching modes, and the less known far--IR bending modes are
the only way to detect these species.

Before the launch of ISO (Kessler et al. 1996), we proposed to observe
the bending modes of several polyatomic molecules in the far--IR.
As a result, the $\nu_2$ bending mode of C$_3$ has been observed
in Sgr~B2 and IRC+10216 (Cernicharo et al.  2000).
In addition, we have detected several Unidentified far-IR bands (UfIBs) in
the ISO spectrum of many C--rich sources that may also be related
with the bending modes of carbon chains species. In particular
we have tentatively assigned a  UfIB at $\sim$57.5~$\mu$m
($\sim$174~cm$^{-1}$), observed  in Sgr~B2 and 
in $C$--rich evolved stars including the planetary nebula (PNe) NGC~7027,
to the $\nu_5$ bending mode of C$_4$ (Cernicharo, Goicoechea \& Benilan 2002).

NGC~7027 is one of the most studied evolved objects in the Galaxy.
It is a young PNe, $\sim$1000~years since it left the AGB stage,
characterized by a relatively compact ionized region (Volk \& Kwok 1997)
driven by the radiation that arises from a central source at 
$\sim$200,000~K (Latter et al. 2000). 
In addition, a rich $C$--chemistry is taking place in the larger molecular
envelope  ($\sim$40$''$ in CO; Masson et al. 1985) that surrounds
the inner photon dominated regions. In fact, NGC~7027 was the first object
to show UIB emission latter attributed to the PAH emission 
(Gillett et al. 1973) and the only source where pure rotational lines
of CH$^+$ have been detected (Cernicharo et al. 1997).

Therefore, the detection of unidentified features in the
IR spectrum of NGC~7027 has historically contributed to a better
understanding of the interstellar carbon complexity.
In this work, we report the detection of a new UfIB at
$\sim$98~$\mu$m ($\sim$102~cm$^{-1}$) in the far--IR spectrum
of NGC~7027 and speculate about it possible origin.

\section{Observations}

The possible emission/absorption produced by the low--energy bending modes of
C$_n$ species has been searched using the \textit{Long--Wavelength
Spectrometer} (LWS; Clegg et al. Swinyard et al. 1996).
In the case of the NGC~7027 PNe, we have used all the
LWS AOT L01 data taken by ISO (see Herpin et al. 2002).
The LWS grating spectra of NGC~7027 taken during orbits
21, 342, 349, 356, 363, 377, 537, 552, 559, 566, 579, 587, 594,
601, 706, 713, 720, 727, 734, 741, 755, 762, 769, 776, and 783 have
been averaged.
The total on-source time was 53409~s and the signal--to--noise (S/N) ratio
is very  high. Here we present part of the spectrum taken with the LWS
between $\sim$86 and 110~$\mu$m ($\sim$116--91~cm$^{-1}$) at
a resolution of $\lambda/\Delta\lambda\simeq200$.
The resulting spectrum  is shown in Fig.~1, \textit{middle panel}.
We also present the same wavelength range in the LWS spectrum
[TDT19800158] of IRC+10216 (see Fig.~1, \textit{lower panel}). The full spectrum 
was shown and modeled by Cernicharo et al. 1996.

All data were processed following
pipeline number 9 and analyzed  using the ISO Spectral Analysis
Package$^2$ (ISAP). Typical routines include: deglitching spikes due
to cosmic rays, oversampling and averaging individual scans and
removing baselines polynomials.
 
\footnotetext[2]
{The ISO Spectral Analysis Package (ISAP) is a joint development by the LWS 
and SWS Instruments Teams and Data Centers. Contributing institutes are CESR, IAS,
 IPAC,MPE, RAL, and SRON.}

In addition, we present part of the  \textit{Short--Wavelength
Spectrometer} (SWS; de Graauw et al. 1996) spectrum of NGC~7027 between
$\sim$4--7~$\mu$m ($\sim$2500-1430~cm$^{-1}$).
The resolution of the SWS01 observation template in this interval is
$\lambda/\Delta\lambda\simeq1500$.
Many carbon chains (polar and non-polar) possess prominent IR active
stretching modes in this wavelength range.
However, Fig.~2 shows that the bulk of the emission
in NGC~7027 arises from narrow ionic and recombination lines from the
inner \Hii region. Only the well--known UIB  broad emission at
$\sim$6.2~$\mu$m
is detected.
This band corresponds to the relaxation of excited vibrational states
of aromatic species pumped by absorption of visible--UV
photons from the hot central star.
The full SWS spectrum has been presented by Bernard Salas et al. (2001).

\section{Results}

After identified and modeled the  molecular rotational
lines of CO $J$=30--29 to 25--24, CH$^+$ $J$=4--3 and OH ${^2}\Pi_{1/2}$
$J$=5/2--3/2 (see Herpin et al. 2002 for the details), a
strong band--like feature constituted by several lines
was identified between 91 and 102~$\mu$m.
Only the high--$J$ CO rotational emission produces modest
contribution in the band wavelengths. Fig.~1 also shows the
maximum contribution expected for the rotational $^{2}\Pi_{1/2}$
5/2--3/2 line of OH at $\sim$98~$\mu$m estimated from the clearer
detections at $\sim$79 and $\sim$84~$\mu$m. Note that no water
lines have been detected in the far--IR spectrum of NGC~7027
(Cernicharo et al. 1997), and thus,
no more $O$-bearing species are contaminating the $\sim$86--110~$\mu$m
window.

Due to the presence of an \Hii region in the inner envelope, we have
also searched for all the possible atomic
transitions that could arise from the nebular gas.
The most likely atomic lines in a PNe such as NGC~7027
(fine structure and recombination lines), taken from the
\textit{Atomic Line List} by Peter van Hoof have been analyzed.
We note that lines from [\Viv], [\Viii] and [\Tiiii],
with transitions in the considered wavelength range, have been previously
detected in the optical spectrum of NGC~7027 (Baluteau et al. 1995),
while several \Hi recombination lines are observed in the mid--IR
(e.g. see Fig.~2).

To estimate the \Hi emission in the far--IR we have
considered  a \Hii region ($2^{''}-4^{''}$) of
$\sim$0.022~\Msun (see the model of Volk \& Kwok 1997)
with  $n_H$=$n_e$ and assumed that the \Hi population is given
by the Saha--Boltzmann equation.
With these parameters, we can predict the \Hi recombination
lines observed in the mid--IR by Bernard Salas et al. (2001),
and estimate the opacity of the far--IR lines. We found that
only the H$n$$\alpha$ series with $n$ from 10 to 15 may produce
modest emission in the LWS range. However, none of these lines
appear between 90 and 110~$\mu$m.
In addition, we have analyzed the possible overlapping with
fine structure lines. In particular, the
[\Tiii]96.68, 106.27  and [\Vii]97.79~$\mu$m lines
could produce some emission in the considered range.
Even assuming a vanadium abundance 10 times larger than the
solar abundance, the [\Vii]97.79~$\mu$m ($^{5}$D$_3$-$^{5}$D$_2$) line would be
extremely weak.
Note that the [\Vii]141.68~$\mu$m ($^{5}$D$_2$-$^{5}$D$_1$) line  is
neither detected in NGC~7027.
Finally, the solar abundance of Ti is $\sim$10 times that
of vanadium. We also computed that is too low to produce significant
emission in the [\Tiii]96.68, 106.27 lines.

Hence, the  integrated band intensity and the different lines
or sub-bands could not
be fully assigned to any of these atomic lines nor to
the pure rotational lines with significant line
strength arising from the light species
in the circumstellar envelope.

We also note that after subtracting the crowed emission of CO, HCN and HCN$_{vib}$
(pure rotational lines in vibrational excited states) from the spectrum of 
IRC+10216, some of the remaining peaks
also agreed with those unidentified peaks observed in NGC~7027
(see Fig. 1). However, the spectral resolution is too poor
to distinguish unidentified features from the forest of
HCN$_{vib}$ rotational lines that contaminate the far--IR spectrum
of IRC+10216 but are missing in NGC~7027.
The HCN abundance in IRC+10216 is very high, $\sim$3$\times$10$^5$, 
with HCN/CO=1/10. In fact, HCN is the main coolant of this $C$--rich
AGB star (Cernicharo et al. 1999). However, the molecule is
photodissociated during the post--AGB evolution as the UV radiation field from 
the evolving star increases. From far--IR observations of HCN pure rotational
lines, Herpin et al. (2002) found minimum abundances of HCN/CO=1/100 and 1/1000
for the proto--PNe CRL~2688 and CRL~618 respectively. 
The extreme case is NGC~7027, were a great part of 
HCN molecules must have been photodissociatd in CN and H.
In fact, Bachiller et al. (1997) found CN/HCN$\sim$10 for this object. Thus, 
the low abundance of HCN compared to IRC+10216 is consistent
with the non detection of far--IR HCN lines (Herpin et al. 2002).
Pure rotational lines of other cyanopolyynes species such as HC$_3$N
can not produce emission in this range because of the high energy
levels associated with its far--IR transitions.

Taking into account the high S/N
ratio of the NGC~7027 spectrum, and the absence of the band
in sources such as Orion, Sgr~A or $O$--rich evolved stars, we believe
that the lines are real.
Due to the chemistry of NGC~7027,
the most probable carrier should be the low--energy bending mode of a
polyatomic molecule containing carbon.
                   
\section{Discussion}

Both PAHs and pure carbon chains possesses low--energy bending modes in the
far--IR. Contrary to the UIB emission in the mid--IR, the far--IR skeletal modes 
of PAHs (vibrations associated with the bending of the skeletal structure)
depends of the exact nature of the species. 
Hence, the rovibrational structure of the PAH skeletal mode has to be calculated 
for each specific PAH. As a example, the lowest energy active
mode of coronene (C$_{24}$H$_{12}$) lies at $\sim$127~cm$^{-1}$
(Joblin et al. 2002), while for ovalene 
(C$_{32}$H$_{14}$), two far--IR skeletal modes at $\sim$120~cm$^{-1}$  
and $\sim$66~cm$^{-1}$ have been theoretically investigated 
(Mulas et al. 2003). The far-IR emission of a few PAHs has also
been observed in gas phase experiments (Zhang et al. 1996).
However, if one of the observed UfIBs belongs to a specific PAHs, other 
bands  (generally stronger) should be observed in the IR spectrum. 
This is not the case of the new
UfIB presented in this work and we have considered the low--energy modes
of pure carbon chains as possible carriers of the observed UfIB.

The detection of several rovibrational lines of the $\nu_2$
bending mode of C$_3$ ($\sim$63~cm$^{-1}$) in Sgr~B2 and
IRC+10216 evidenced  high abundances for this species
and opened the possibility to detect more  carbon chains
in the ISM and CSM (Cernicharo et al. 2000).
At the low resolution of the LWS/grating, the more intense
C$_3$ lines [Q(2,4,6)] are blended with the strong
[\CII]158~$\mu$m line and no assignation could be made
in NGC~7027.
However, the tentative detection of the $\nu_5$ bending
mode of C$_4$ in many objects including NGC~7027
($\sim$174~cm$^{-1}$; Cernicharo et al. 2002),
if confirmed, will imply a C$_3$/C$_4$$<$10
abundance ratio, which will suggest that the abundance of C$_n$ does not
decrease  drastically when $n$ increases.

In spite of their importance, the C$_n$  bending modes are difficult
to characterize spectroscopically in the laboratory or
by \textit{ab initio} computations.
As expected, the main uncertainties difficulting their astronomical
detection are the band origin and the IR intensity.
Moreover, the induced dipole moments
and other spectroscopic constants
are often  known for the fundamental transition but less or nothing
is known about overtone transitions within the bending mode or transitions
to other excited states including the stretching modes.

The ground electronic state is  different
for the odd-- and even-- numbered cumulenic chains. This property determines
the observed band shape of the bending mode.
Odd--numbered linear C$_n$'s have a singlet
$^{1}\Sigma^{+}_{g}$ ground electronic state which results in
a perpendicular $^{1}\Pi_{vib}-{^1}\Sigma_{vib}$ vibronic spectrum with a
strong --$Q$ branch and weak --$P$ and --$R$ branches.
On the other hand, even--numbered linear C$_n$'s have a triplet
$^{3}\Sigma^{-}_{g}$ ground electronic state. Hence, each rotational
line  is split in three components
due to  spin couplings. This splitting
increase with the number of atoms but is  particularly
small for C$_4$ and C$_6$ (Giesen et al. 2001).
The vibronic transition is now $^{3}\Pi_{vib}-{^3}\Sigma_{vib}$, and
despite the null component of the electronic orbital
momentum in the ground state, the spin--orbit constant (A$_{SO}$)
could be large in  even-- C$_n$'s.
In such a case, the resulting band--shape  will
be notoriously affected by the value of A$_{SO}$.

\subsection{Tentative detection of C$_6$ and C$_5$}

With the exception of the $\nu_2$ mode of C$_3$, the band origins of
other C$_n$ bending modes are not accurately constrained.
Hence, their assignation as carriers of the UfIBs
is not obvious. Although the low--lying bending modes of  C$_5$ and C$_6$
have not been directly observed in gas phase, different studies suggest wavenumbers
around $\sim$100~cm$^{-1}$ (see below).
The observed new UfIB around $\sim$98~$\mu$m
is composed by several peaks at
$\sim$91.8, 92.8,  93.9, 95.2, 97.6, 98.9, 100.5 and 101.7~$\mu$m.
In the following, we will consider the possibility that
 \textit{any of these peaks is  related with the
$\nu_9$ and $\nu_7$ bending modes of C$_6$ and C$_5$ respectively}.

C$_5$ was first detected in the gas phase by Heath et al. (1989)
who observed the $\nu_3$ stretching mode at $\sim$2169~cm$^{-1}$
($\sim$4.6~$\mu$m). The same mode was observed in
IRC+10216 by Bernath et al. (1989). Actually,  C$_5$ is the largest
C$_n$ detected in the CSM.
On the other hand, C$_6$ was first identified  in the laboratory by its
electronic spin resonance spectrum in an Ar matrix (Van Zee et al. 1987)
and later observed in gas phase through its $\nu_4$ stretching mode at
$\sim$1960~cm$^{-1}$ ($\sim$5.1~$\mu$m) by Hwang et al. (1993).
Moazzen--Ahmadi et al.
(1989) observed the ($\nu_3$+$\nu_7$)--$\nu_7$ and ($\nu_3$+2$\nu_7$)--2$\nu_7$
hot bands arising from the $\nu_7$ bending mode of C$_5$.
From the $l$--doubling constant $q_7$, they estimated a frequency of
$\nu_7$=118$\pm$3~cm$^{-1}$.
More recently, C$_5$ has been observed in photoelectron spectra.
In particular, Arnold et al. (1991) measured the 2$\nu_7$ transition
and estimated $\nu_7$=101$\pm$45~cm$^{-1}$, while Kitsopoulos et al.
(1991) obtained $\nu_7$=107$\pm$5~cm$^{-1}$. \textit{Ab initio}
calculations predict an infrared intensity around  36~km~mol$^{-1}$
[$A$($\nu_7$=1-0)$\simeq$0.06~s$^{-1}$; Hutter \& L\"uthi  1994; Martin et al.
1995]. Fig.~1 shows the expected band--shape for a
$^{1}\Pi_{vib}-{^1}\Sigma_{vib}$ transition with this intensity and  the molecular
constants for the $\nu_7$ mode (Moazzen--Ahmadi et al.)
centered at 104.8~cm$^{-1}$ (95.4~$\mu$m). As said before, any
of the band peaks observed in NGC~7027 around
$\sim$98~$\mu$m ($\sim$102~cm$^{-1}$) could well be responsible
of the $\nu_7$ mode. However, only the best fit to the
band including the C$_6$ model
is shown in  Fig.~1.
If the carrier of the 95.5~$\mu$m feature is finally C$_5$
at $\nu_7$=104.8~cm$^{-1}$,
the estimated column density
assuming an excitation temperature 
of 100~K (typical of a PNe envelope) is
$N$(C$_5$)=1.8$\times$10$^{14}$~cm$^{-2}$.
Assuming 1.5--4.0  magnitudes of visual extinction in the neutral envelope
(Hasegawa et al. 2000), the typical abundance of C$_5$ in NGC~7027
would be (0.5--1.0)$\times$10$^{-7}$, similar to that derived by Bernath
et al. (1989) in IRC+10216.

Fig.~2 shows that, for the same column densities required to
reproduce the bending modes, the $\nu_4$ and $\nu_3$ stretching
modes of C$_6$ and C$_5$ are not detected in NGC~7027.
Different excitation conditions of the stretching and bending
modes and geometrical effects may explain this conjecture. 
Although carbon clusters have a moderate
size, we could also expect an efficient pumping of the stretching and
bending modes through UV photons. Mid--IR photons could also contribute
to the excitation of the high--$\nu_{bending}$
excited states but also of the stretching modes. Finally, the low--energy bending
modes could be pumped alone by the absorption of far--IR dust photons
(see the case of the C$_3$ excitation in  Cernicharo et al. 2000).
However, the smaller volume of the photon--dominated regions relative to
the far--IR dusty envelope in NGC~7027, and the larger mid--IR flux of
IRC+10216 (where the C$_3$ and C$_5$ stretching modes were detected)
compared to NGC~7027 (a factor $\sim$10$^3$ at 5~$\mu$m, but only a factor
$\sim$4 at 100~$\mu$m), can favor the observation of the C$_n$ bending 
modes within the large ISO beam.
Under these conditions, the detection of the bending modes and the
non--detection of the stretching modes in NGC~7027 can be plausible.\\

The $\nu_9$ low--energy bending mode of  C$_6$ is even less known.
\textit{Ab inito} calculations predict a $\nu_9$$\sim$108~cm$^{-1}$
frequency and an infrared intensity of $\sim$25~km~mol$^{-1}$
[$A$($\nu_9=1-0$)$\simeq$0.035~s$^{-1}$; Martin et al. 1990; 1995), while
photoelectron spectroscopy  experiments estimate
$\nu_9$=97$\pm$45~cm$^{-1}$ (Arnold et al. 1991) and 
$\nu_9$=90$\pm$50~cm$^{-1}$  (Xu et al. 1997).
Theoretical predictions have to be taken into account with caution
since they rely in the harmonic approximation which can be in error
for large amplitude benders such as radicals with low-lying
electronic states. In particular, the infrared intensity has to be
considered as a lower limit (and column densities as upper limits).
\textit{Ab initio}  calculations do predict a low-lying electronic
state for open--shell species such as C$_4$ or C$_6$. The term value for the
$^{1}\Delta_g$ lowest energy excited state of C$_6$ has been recently
determined by Xu et al. only at  $\sim$1400~cm$^{-1}$  from the ground.
Hence, A$_{SO}$  can effectively be large even in the ground
state ($\Lambda$=0).
As an example,
the A$_{SO}$ constant of C$_4$H ($^{2}\Sigma$ ground state) is
very large, $\simeq$3~cm$^{-1}$ (Yamamoto et al. 1987), because its
$^{2}\Pi$ electronic excited state is only $\simeq$468~cm$^{-1}$
above the ground. In the case of even--numbered C$_n$ chains,
A$_{SO}$ can easily be larger because of the higher  spin multiplicity
of the ground state and the larger $\Lambda$ value
of the lowest--lying electronic excited state.
Fig.~1 also shows the expected band--shape for a
$^{3}\Pi_{vib}-{^3}\Sigma_{vib}$ transition with A$_{SO}$=3.5~cm$^{-1}$
(we estimated A$_{SO}$$\sim$4~cm$^{-1}$ for C$_4$),
C$_6$  molecular constants  from  Hwang et al. (1993) and van
Zee et al. (1987;1988), and  the $\nu_9$ band origin at
98.3~cm$^{-1}$ (101.7~$\mu$m). Assuming the same excitation temperature
than for C$_5$ we derive $N$(C$_6$)=0.8$\times$10$^{14}$~cm$^{-2}$.
Hence, the C$_6$ abundance will be  a factor 2 smaller
than that of C$_5$.

Much more  work has to be done to fully understand and characterize
the low--energy vibrations of C$_n$ chains in order to assign the UfIBs
observed by ISO. The  cyclic C$_n$ isomers could also produce
spectral features in the far--IR, however, their active modes are even
less known.
Waiting for such progresses, space observations offer the unique 
opportunity to obtain spectra in the far--IR domain where the bending 
modes appear and to motivate further laboratory and theoretical studies.
The presence of small C$_n$ chains  in the space
as well as their high reactivity, suggest that these species can be
involved in the formation of more complex organic molecules.
In fact, for cumulenic clusters with $n$=10 to 20, linear structures
are thought to close into rings, while for  $n$$>$30,
these species are thought to be more stable in aromatic and fullerene--like
structures (O' Brien et al. 1987).

Future space heterodyne telescopes, such as the \textit{Herschel Space
Observatory}, with much better sensitivity and
spectral resolution in the far--IR,
should allow the detection of longer C$_n$ chains through their
low--energy bending modes. This will be the fingerprint needed to
understand  the formation and the nature of the UIBs and UfIBs carriers.

\clearpage

\acknowledgments

We thank Spanish DGES and PNIE for funding
support under grants PANAYA2000-1784, ESP2001-4516, AYA2002-10113-E,
ESP2002-01627, AYA2002--02117  and AYA2003-02785-E.
We also thank Ana Heras for providing us the processed SWS
spectrum of NGC~7027 and F.~Najarro and C.~Joblin for useful comments about
atomic lines and PAHs respectively. We have used the Atomic Line List v2.04
of Peter van Hoof (http://www.pa.uky.edu/ $\sim$peter/atomic).

\begin{figure}
\caption
{\textit{Top panel}:
Expected band shape for a $^1\Pi_{vib}-{^1}\Sigma_{vib}$ transition
(as the $\nu_7$ mode of C$_5$)  and
for a $^3\Pi_{vib}-{^3}\Sigma_{vib}$ transition (as the $\nu_9$ mode of C$_6$)
with A$_{SO}$=3.5~cm$^{-1}$. The excitation temperature for both
transitions is 100~K. The intrinsic line width is 10~km~s$^{-1}$
and the spectral resolution $\lambda/\Delta\lambda$ is 200.
---\textit{Middle panel}: Observed ISO/LWS spectrum of NGC~7027
between $\sim$86 and 110~$\mu$m. The ordinate corresponds to the
flux over the  continuum flux and the abscissa to the wavelength in $\mu$m.
Large Velocity Gradient model for the pure
rotational emission of CO, CH$^+$ and OH (Herpin et al. 2002) and
the total model for the  bending modes of
C$_5$ and  C$_6$.
The thick horizontal arrow represents
the observational uncertainty for the $\nu_7$ C$_5$
band origin (from Kitsopoulos et al. 1991). The experimental
uncertainty for the $\nu_9$ C$_6$ mode is 90$\pm$50~cm$^{-1}$
(Xu et al. 1997). See text.
---\textit{Bottom panel}:  Observed ISO/LWS spectrum of IRC+10216
between $\sim$86 and 110~$\mu$m. The ordinate corresponds to the
continuum subtracted flux  and the abscissa to the wavelength in $\mu$m.
Large Velocity Gradient model for the pure
rotational emission of CO $\nu$=0, 1, $^{13}$CO $\nu$=0,  HCN and H$^{13}$CN
$\nu=$0, and HCN$_{vib}$ $\nu_2$=1, 2 and $\nu_{1,3}$=1 (Cernicharo et al. 1996).
The rotational transitions of CO, HCN in the ground state, and
HCN$_{vib}$ $\nu_2$=1, 2 and $\nu_{1,3}$=1 are indicated.
}
\end{figure}
\plotone{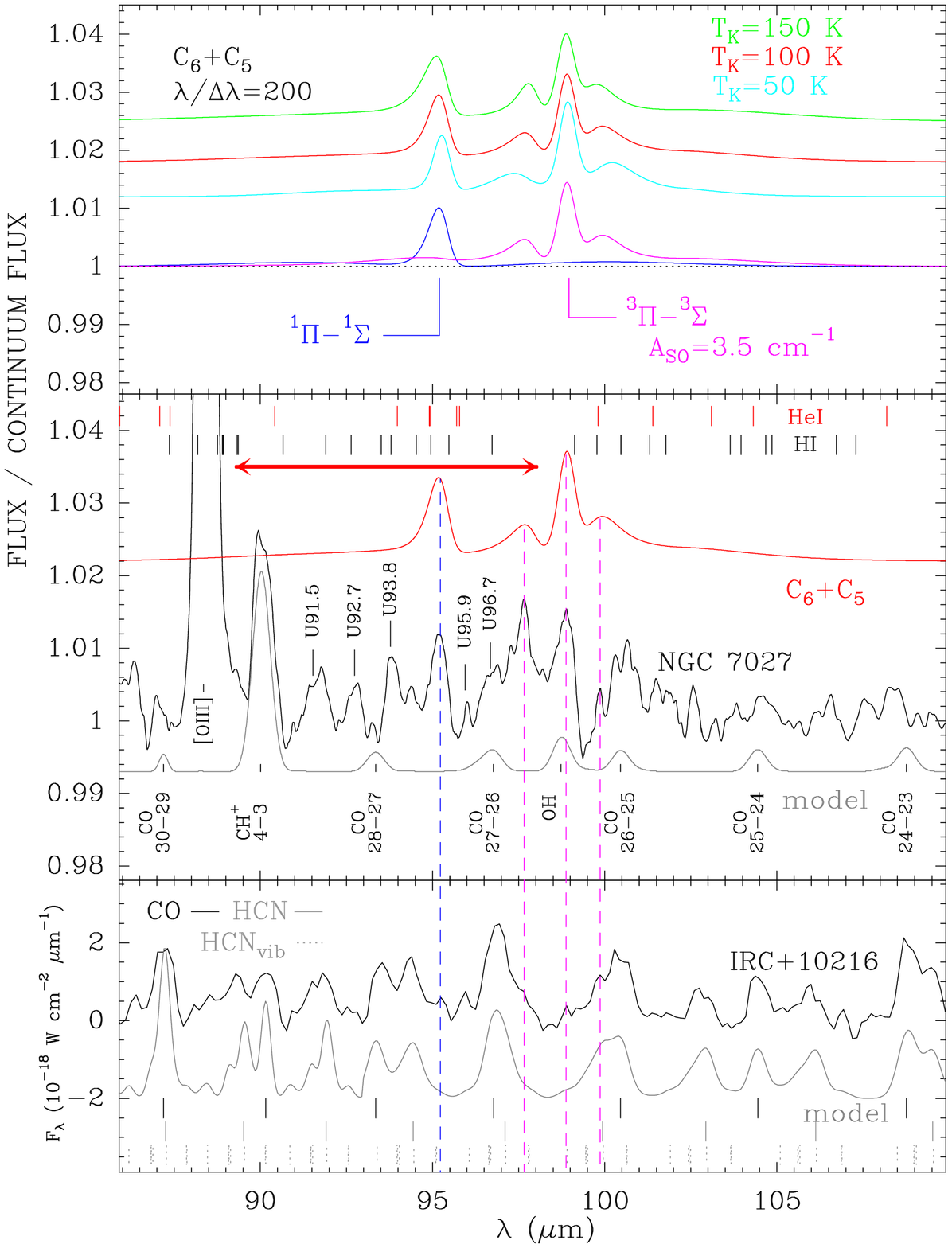}

\clearpage

\begin{figure}
\caption{Observed ISO/SWS spectrum of NGC~7027
between $\sim$4 and 7~$\mu$m. The ordinate scale is F/F$_c$ and the
wavelength is in $\mu$m. The fine structure and recombination
lines from the ionized region are labeled.
The expected $\Sigma_{vib}-\Sigma_{vib}$ parallel spectra from the
$\nu_3$ C$_3$, $\nu_3$ C$_4$, $\nu_3$ C$_5$, $\nu_4$ C$_6$
and $\nu_4$ C$_7$  stretching modes are
also shown. The excitation temperature for all bands is also
100~K. The intrinsic line width is 10~km~s$^{-1}$
and the spectral resolution $\lambda/\Delta\lambda$ is 1500.
Column densities are $N$(C$_5$)=1.8$\times$10$^{14}$~cm$^{-2}$
(from the $\nu_7$ bending mode), $N$(C$_6$)=0.8$\times$10$^{14}$~cm$^{-2}$
(from the $\nu_9$ bending mode), $N$(C$_3$)=10$\times$$N$(C$_5$),
$N$(C$_7$)=$N$(C$_5$)$/$10 and $N$(C$_4$)=0.8$\times$10$^{15}$~cm$^{-2}$
(from Cernicharo et al. 2002).}
\end{figure}
\plotone{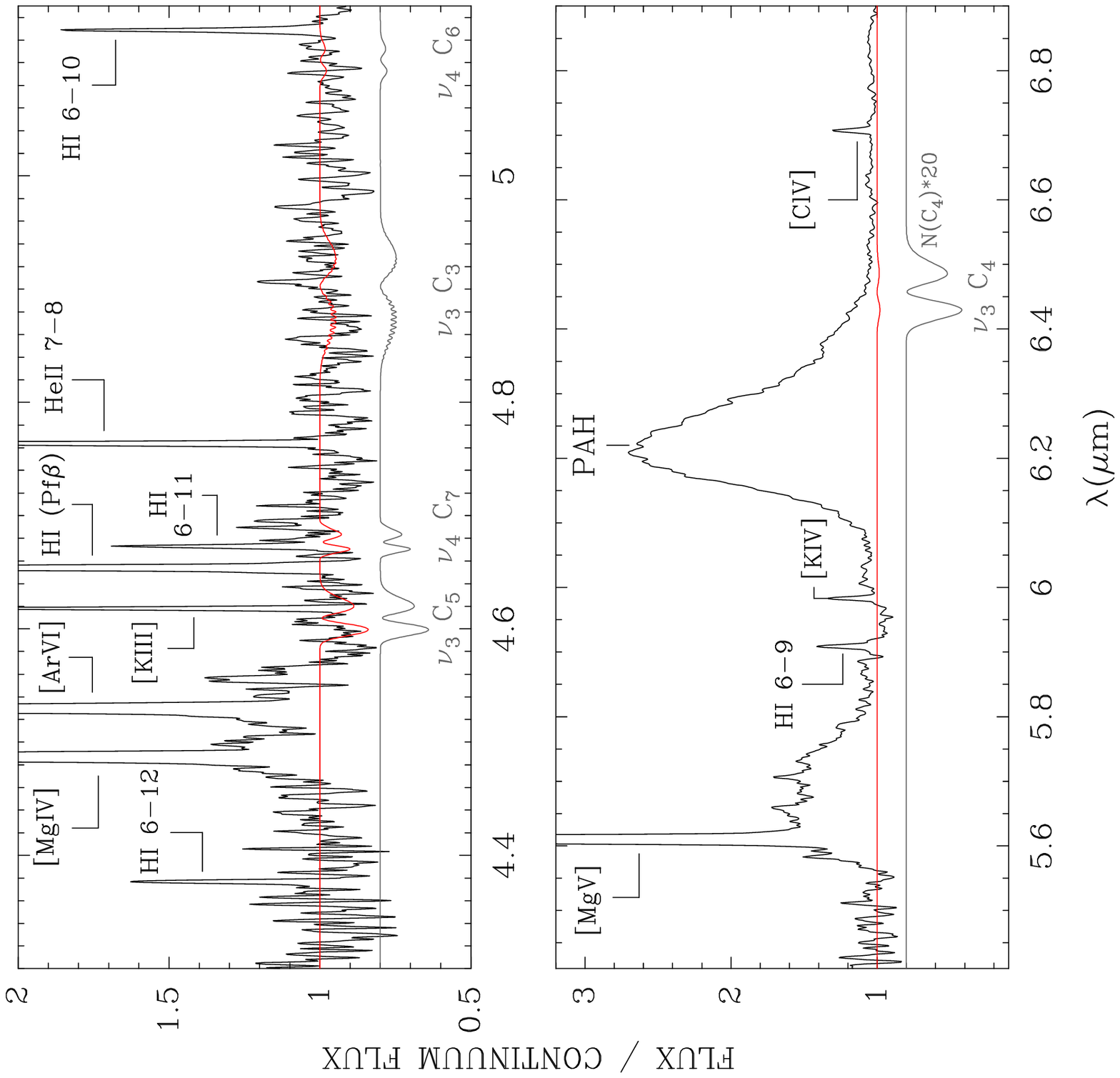}

\end{document}